\documentclass[twocolumn,english,reprint,preprintnumbers,amsmath,amssymb,aps]{revtex4}
\usepackage[T1]{fontenc}
\usepackage[latin9]{inputenc}
\usepackage{textcomp}
\usepackage{amsmath}
\usepackage{amssymb}
\usepackage{esint}

\makeatletter

\providecommand{\tabularnewline}{\\}

\@ifundefined{textcolor}{}
{%
 \definecolor{BLACK}{gray}{0}
 \definecolor{WHITE}{gray}{1}
 \definecolor{RED}{rgb}{1,0,0}
 \definecolor{GREEN}{rgb}{0,1,0}
 \definecolor{BLUE}{rgb}{0,0,1}
 \definecolor{CYAN}{cmyk}{1,0,0,0}
 \definecolor{MAGENTA}{cmyk}{0,1,0,0}
 \definecolor{YELLOW}{cmyk}{0,0,1,0}
}

\usepackage{dcolumn}
\usepackage{bm}
\usepackage{amsfonts}\usepackage{babel}
\providecommand{\U}[1]{\protect\rule{.1in}{.1in}}

\@ifundefined{textcolor}{}{
\definecolor{BLACK}{gray}{0}
 \definecolor{WHITE}{gray}{1}
 \definecolor{RED}{rgb}{1,0,0}
 \definecolor{GREEN}{rgb}{0,1,0}
 \definecolor{BLUE}{rgb}{0,0,1}
 \definecolor{CYAN}{cmyk}{1,0,0,0}
 \definecolor{MAGENTA}{cmyk}{0,1,0,0}
 \definecolor{YELLOW}{cmyk}{0,0,1,0}
 }

\usepackage{babel}
\usepackage{ragged2e}

\usepackage{babel}

\makeatother

\usepackage{babel}
\begin{document}

\title{Superselection rule for the cosmological constant in three-dimensional
spacetime}

\author{Claudio Bunster}

\email{bunster@cecs.cl}

\affiliation{Centro de Estudios Científicos (CECs), Av. Arturo Prat 514, Valdivia,
Chile}

\author{Alfredo Pérez}

\email{aperez@cecs.cl}

\affiliation{Centro de Estudios Científicos (CECs), Av. Arturo Prat 514, Valdivia,
Chile}


\begin{abstract}
Efforts to understand the origin of the cosmological constant $\Lambda$
and its observed value have led to consider it as a dynamical field
rather than as a universal constant. Then the possibility arises that
the universe, or regions of it, might be in a superposition of quantum
states with different values of $\Lambda$, so that its actual value
would not be definite. There appears to be no argument to rule out
this possibility for a generic spacetime dimension $D$. However,
as proved herein, for $D=3$ there exists a superselection rule that
forbids such superpositions. The proof is based on the asymptotic
symmetry algebra. 
\end{abstract}
\maketitle

\section{Introduction}

Superselection rules \cite{WWW} lay at the very foundation of quantum
mechanics. One says that the hermitian operator $\hat{Q}$ obeys a
superselection rule if the Hilbert space $\mathcal{H}$ of the system
is a direct sum of subspaces $\mathcal{H}_{j}$ belonging to different
eigenvalues of $\hat{Q}$, 
\[
\mathcal{H}=\bigoplus_{j}\mathcal{H}_{j}.
\]
The $\mathcal{H}_{j}$ are called coherent subspaces or superselection
sectors. In each of them the superposition principle is valid, but
a linear combination, 
\[
\alpha\psi_{1}+\beta\psi_{2},
\]
of states $\psi_{1}$ and $\psi_{2}$ from two distinct coherent subspaces
is not physically realizable, except as a mixture with the density
matrix, 
\[
\left|\alpha\right|^{2}\psi_{1}\otimes\psi_{1}^{\dagger}+\left|\beta\right|^{2}\psi_{2}\otimes\psi_{2}^{\dagger}.
\]
An alternative equivalent description is that the relative phase $\beta$
and $\alpha$ in $\alpha\psi_{1}+\beta\psi_{2}$ is not observable.

If $\hat{Q}$ obeys a superselection rule, then not every hermitian
operator is observable, but only those with commute with $\hat{Q}$.
This is because an eigenvector of an operator not commuting with $\hat{Q}$
would be a superposition of different eigenvectors of $\hat{Q}$.

Since the introduction of the concept in \cite{WWW}, superselection
rules, such as the one for electric charge, have not been devoid of
controversy (see for example \cite{W}). Cases in which a superselection
rule can be distinctly proven are somewhat scarce and precious. The
proof is then often based on a symmetry principle, such as it is the
case for the fermion-boson superselection rule and for the Bargmann
superselection rule for the mass in non-relativistic quantum mechanics,
which will be recalled below.

Spacetime symmetry-albeit asymptotic- will also be at the heart of
the superselection rule for the cosmological constant in three-dimensional
spacetime, which is the main new result reported in this paper.

The plan of the paper is the following: Section \ref{sec:SRQM} recalls
in some detail the superselection rule for the mass in non-relativistic
quantum mechanics. This is done because the parallel between the concepts
occurring in that simple case, and those appearing in the more involved
situation of the cosmological constant, is an extremely close one.
Section \ref{sec:SRG} is then devoted to the superselection for the
cosmological constant, the proof of which is based on the asymptotic
symmetry algebra of three-dimensional gravity with the cosmological
constant $\Lambda$ treated as a dynamical variable. This algebra
turns out to be the same as the one found in \cite{BH} for a fixed
non-dynamical $\Lambda$.

The analysis of the asymptotic algebra when $\Lambda$ is dynamical
is not devoid of subtleties, which are of the same nature of those
previously found in four-dimensional spacetime \cite{HTAdS}. In order
not to interrupt the thread of the discussion of the superselection
rule those subtleties are dealt with in the Appendix.

\begin{table*}
\protect\protect\caption{\label{tab:table3}Superselection rules for the mass in non-relativistic
quantum mechanics and for the cosmological constant in three-dimensional
spacetime compared and contrasted}

\begin{ruledtabular} %
\begin{tabular}{ccc}
 & \multicolumn{1}{c}{Non-relativistic quantum mechanics} & \multicolumn{1}{c}{Three-dimensional gravity}\tabularnewline
\hline 
Spacetime symmetry algebra  & Galilei algebra  & Two Witt algebras \tabularnewline
Centrally extended algebra  & Bargmann algebra  & Two Virasoro algebras \tabularnewline
Central charge  & $m$  & $3\ell/2G$\tabularnewline
Superselected quantity  & $m$  & $\Lambda=-1/\ell^{2}$\tabularnewline
\end{tabular}\end{ruledtabular}
\end{table*}

\section{Superselection rule for the mass in non-relativistic quantum mechanics\label{sec:SRQM}}

\subsection{Galilei Lie algebra }

The superselection rule for the mass in non-relativistic quantum mechanics
appears as a consequence of the transformation properties of the states
under the Galilei algebra.

The Galilei group is characterized by the following set of transformations
acting on space and time, 
\begin{align}
x'_{i} & =R_{ij}x_{j}+v_{i}t+a_{i},\label{eq:galilei1}\\
t' & =t+b,\label{eq:Galilei2}
\end{align}
where $R_{ij}=R_{ji}$ characterize the spatial rotations, $v_{i}$
the Galilei boosts and $a_{i}$ and $b$ translations in the space
and the time respectively. From eqs. (\ref{eq:galilei1}) and (\ref{eq:Galilei2})
one obtains directly the Lie algebra of the Galilei group when it
acts on points $x_{i},t$ of spacetime, 
\begin{align*}
\left[J_{i},J_{j}\right] & =\epsilon_{ijk}J_{k}\;,\;\;\;\left[J_{i},P_{j}\right]=\epsilon_{ijk}P_{k}\;,\;\;\;\left[J_{i},K_{j}\right]=\epsilon_{ijk}K_{k},
\end{align*}
\begin{align}
\left[K_{i},H\right]=P_{i} & \;,\;\;\;\left[P_{i},P_{j}\right]=0\;,\;\;\;\left[K_{i},K_{j}\right]=0,\label{eq:galileiwithoutc}
\end{align}
\begin{align*}
\left[J_{i},H\right]=0 & \;,\;\;\;\left[P_{i},H\right]=0\;,\;\;\;\left[K_{i},P_{j}\right]=0.
\end{align*}
Here $H,\vec{P},\vec{J}$ and $\vec{K}$ are the generators of time
translations, space translations, rotations and boosts respectively.
The above algebra possesses no central charge.

\subsection{Canonical realization of the Galilei group algebra for a free particle}

If $\vec{x}$ and $\vec{p}$ respectively denote the position and
its conjugate momentum, the generators are realized as, 
\begin{align*}
H & =\frac{\vec{p}^{2}}{2m},\;\;\vec{P}=\vec{p},\;\;\vec{J}=\vec{x}\times\vec{p},\;\;\vec{K}=m\vec{x}.
\end{align*}
In terms of Poisson brackets, these generators obey, 
\[
\left\{ J_{i},J_{j}\right\} =\epsilon_{ijk}J_{k},\;\:\left\{ J_{i},P_{j}\right\} =\epsilon_{ijk}P_{k},\;\:\left\{ J_{i},K_{j}\right\} =\epsilon_{ijk}K_{k},
\]
\begin{align}
\left\{ K_{i},H\right\} =P_{i} & \;,\;\;\;\left\{ P_{i},P_{j}\right\} =0\;,\;\;\;\left\{ K_{i},K_{j}\right\} =0,\label{eq:galileiwithoutc-1-1}
\end{align}
\begin{align*}
\left\{ J_{i},H\right\} =0 & \;,\;\;\;\left\{ P_{i},H\right\} =0\;,\;\;\;\left\{ K_{i},P_{j}\right\} =m\delta_{ij}.
\end{align*}
We see that the canonical realization of the algebra acquires a central
charge.

\subsection{Quantum mechanics and superselection rule}

In passing to quantum mechanics one retains the algebra (\ref{eq:galileiwithoutc-1-1})
in terms of operators by replacing the Poisson brackets by the commutator
divided by $i\hbar$. Additionally one promotes the central charge
to be a hermitian operator $\hat{m}$ which becomes a new generator
in a ``centrally extended'' algebra %
\footnote{\label{fn:At-the-classical}At the classical level the central charge
in the Poisson bracket algebra (\ref{eq:galileiwithoutc-1-1}) may
be turned into a dynamical variable by introducing an additional canonical
pair $m\left(t\right),\tau\left(t\right)$ such that the action reads,
$I=\int dt\left(\vec{p}\cdot\dot{\vec{x}}+m\dot{\tau}-\frac{\vec{p}^{2}}{2m}\right)$.
See \cite{Giulini} for elaborations on this simple procedure.%
}. This yields the Bargmann algebra \cite{Bargmann},

\begin{align*}
\left[\hat{J}_{i},\hat{J}_{j}\right] & =i\hbar\epsilon_{ijk}\hat{J}_{k},\;\;\;\left[\hat{J}_{i},\hat{P}_{j}\right]=i\hbar\epsilon_{ijk}\hat{P}_{k},
\end{align*}
\[
\left[\hat{J}_{i},\hat{K}_{j}\right]=i\hbar\epsilon_{ijk}\hat{K}_{k},\;\;\;\left[\hat{K}_{i},\hat{H}\right]=i\hbar\hat{P}_{i},
\]
\[
\left[\hat{P}_{i},\hat{P}_{j}\right]=0,\;\;\;\left[\hat{K}_{i},\hat{K}_{j}\right]=0,\;\;\;\left[\hat{J}_{i},\hat{H}\right]=0,
\]
\[
\;\;\;\left[\hat{P}_{i},\hat{H}\right]=0,\;\;\;\left[\hat{K}_{i},\hat{P}_{j}\right]=i\hbar\hat{m}\delta_{ij},
\]
\[
\left[\hat{m},\hat{J}_{i}\right]=0,\;\;\;\left[\hat{m},\hat{P}_{i}\right]=0,\;\;\;\left[\hat{m},\hat{K}_{i}\right]=0,\;\;\;\left[\hat{m},\hat{H}\right]=0.
\]

The action of the Bargmann algebra on the quantum states has important
physical consequence. It implies a superselection rule for the mass
in non-relativistic quantum mechanics.

The key point is the following. One assumes that the observations
are realized in spacetime. Therefore one postulates that a sequence
of operations that has no effect on spacetime should have no physical
effect, and hence should at most alter all quantum mechanical states
by a common phase. Now, when the mass is allowed to be a dynamical
variable, the action of the Galilei transformations on quantum mechanical
states does not obey the Galilei algebra, it obeys the Bargmann algebra.
Due to the central extension, a sequence of transformations which
is the identity in the spacetime may alter the relative phase of different
state vectors. This is the origin of the superselection rule.

Thus one considers a sequence of transformations with infinitesimal
parameters $\vec{v},\vec{a},-\vec{v},-\vec{a}$. The action of this
sequence on spacetime is the identity, so it should alter a permissible
state vector $\psi$ by, 
\begin{equation}
\delta\psi=i\theta\psi.\label{eq:theta}
\end{equation}
However, it follows from the Bargmann algebra that,

\[
\delta\psi=\frac{1}{\hbar^{2}}\left[\vec{v}\cdot\hat{\vec{K}},\vec{a}\cdot\hat{\vec{P}}\right]\psi=i\left(\vec{v}\cdot\vec{a}\right)\frac{\hat{m}}{\hbar}\psi,
\]
Now, if one has, 
\begin{equation}
\psi=\psi_{1}+\psi_{2},\label{eq:superp}
\end{equation}
where $\psi_{1}$ and $\psi_{2}$ are two different eigenstates of
the mass operator one obtains, 
\[
\delta\psi=\frac{i}{\hbar}\vec{v}\cdot\vec{a}\left(m_{1}\psi_{1}+m_{2}\psi_{2}\right),
\]
which is not of the form (\ref{eq:theta}). Therefore the superposition
(\ref{eq:superp}) is not allowed in the Hilbert space and the superselection
rule is established.

\section{Superselection rule for the cosmological constant\label{sec:SRG}}

\subsection{Quantum mechanics and superselection rule}

For gravity with a negative cosmological constant in three spacetime
dimensions, the role of the Galilei algebra for the non-relativistic
particle is played by its global symmetry algebra. This algebra is
the asymptotic form at large distances of the algebra of deformations
\cite{T} of a two-dimensional spacelike hypersurface embedded in
the three-dimensional spacetime. It is of infinite dimension \cite{BH}
and it consists of two copies of the Witt algebra: 
\[
\left[L_{m}^{\pm},L_{n}^{\pm}\right]=\left(m-n\right)L_{m+n}^{\pm}.
\]
These equations are the analog of (\ref{eq:galileiwithoutc}).

When the symmetry is canonically realized in terms of Poisson brackets,
the algebra is centrally extended \cite{BH} and each copy of the
Witt algebra becomes the Virasoro algebra, 
\begin{equation}
i\left\{ L_{m}^{\pm},L_{n}^{\pm}\right\} =\left(m-n\right)L_{m+n}^{\pm}+\frac{c}{12}\left(m^{3}-m\right)\delta_{m+n,0}.\label{eq:Vir}
\end{equation}
The central charge $c$ in (\ref{eq:Vir}) is expressed in term of
Anti-de Sitter radius of curvature $\ell$ and the gravitational constant
$G$ by, 
\begin{equation}
c=3\ell/2G.\label{eq:c}
\end{equation}
The radius $\ell$ is related to the cosmological constant by, 
\[
\Lambda=-\frac{1}{\ell^{2}}.
\]

To establish the desired superselection rule, one promotes the central
charge to an operator $\hat{c}$ and retains the algebra (\ref{eq:Vir})
in terms of operators, replacing the Poisson brackets by the commutator
divided by $i\hbar$. 
\begin{align}
\left[\hat{L}_{m}^{\pm},\hat{L}_{n}^{\pm}\right] & =\hbar\left(m-n\right)\hat{L}_{m+n}^{\pm}+\frac{\hbar\hat{c}}{12}\left(m^{3}-m\right)\delta_{m+n,0},\nonumber \\
\left[\hat{L}_{m}^{\pm},\hat{c}\right] & =0,\label{eq:Vir1}\\
\left[\hat{c},\hat{c}\right] & =0.\nonumber 
\end{align}
Next, just as for the particle case, one looks for a sequence of infinitesimal
symmetry operations which will be the identity when the central charge
is turned-off, i.e., when acting on spacetime, but which will give
a non-trivial result when the central charge is turned on.

It will be sufficient to consider one copy of the Virasoro algebra,
and within it, the operators $\hat{L}_{n},\hat{L}_{-n}=\hat{L}_{n}^{\dagger}$
for any fixed $n\geq2$, and $\hat{L}_{0}$. We may write, 
\[
\hat{L}_{n}=\hat{K}_{n}+i\hat{P}_{n},
\]
where $\hat{K}_{n}$ and $\hat{P}_{n}$ are hermitian. It is then
clear from (\ref{eq:Vir1}) that the central charge appears only in
the commutator of $\hat{K}_{n}$ with $\hat{P}_{n}$, which obeys,
\begin{equation}
\left[\hat{K}_{n},\hat{P}_{n}\right]-i\hbar n\hat{L}_{0}=\frac{i\hbar\hat{c}}{24}n\left(n^{2}-1\right).\label{eq:KP}
\end{equation}
When the central charge is turned-off (Witt algebra), the left-hand
side vanishes, which means that the Lie algebra element has no effect
when acting on spacetime, and one should demand that it does not alter
the physical state. If one compares this statement with the corresponding
one for the non-relativistic particle, one sees that the operation
involved is the composition of five elementary operations rather than
four. If one calls $v_{n},a_{n}$ and $b$ the infinitesimal parameters
corresponding to $\hat{K}_{n},\hat{P}_{n}$ and $\hat{L}_{0}$, the
sequence is given by $v_{n},a_{n},-v_{n},-a_{n},b=-nv_{n}a_{n}$.
It follows from (\ref{eq:KP}) that the effect of the resulting transformation
is given by,

\[
\delta\psi=\frac{in\left(n^{2}-1\right)}{24\hbar}\left(v_{n}a_{n}\right)\hat{c}\psi.
\]

Now, if one has, 
\[
\psi=\psi_{1}+\psi_{2},
\]
where $\psi_{1}$ and $\psi_{2}$ are two different eigenstates of
the central charge operator one obtains, 
\begin{align*}
\delta\psi & =\frac{in\left(n^{2}-1\right)}{24\hbar}v_{n}a_{n}\left(c_{1}\psi_{1}+c_{2}\psi_{2}\right)\neq i\theta\psi,
\end{align*}
which proves the superselection rule %
\footnote{The Galilei algebra may be deformed into the Poincaré algebra, of
which it is the contraction when the speed of light goes to infinity.
The superselection rule for the (rest) mass does not hold for the
Poincaré case, which has no central charge (unstable relativistic
particles do exist!). A natural question to ask is whether a similar
phenomenon might occur for the cosmological constant, i.e, whether
it would be possible to ``decontract'' the Virasoro algebra to one
without central charge. The answer is in the negative since the Virasoro
algebra is rigid \cite{Fialowski}. We thank Prof. Marc Henneaux for
informing us of this fact. %
}.

\subsection{Superselection rule is for a dynamical cosmological constant with
a fixed universal gravitational constant }

In the case of the non-relativistic particle one had a finite number
of degrees of freedom and a finite number of symmetry generators in
the Galilei algebra. Since after all the superselection rule is quantum
mechanical, one might as well have skipped the classical mechanics
altogether. It would have been sufficient to simply decree that the
mass $m$ was an additional hermitian operator, commuting with all
the original degrees of freedom and therefore with all the original
symmetry generators. We included the Poisson bracket algebra to emphasize
the close analogy with the three-dimensional gravity case (which shows
\emph{en passant} that ``classical central charges'', i.e. central
charges arising in the canonical realization of spacetime symmetries,
have been around for quite a while %
\footnote{It is curious to realize how many people, including the present authors,
felt since the early days of string theory, that central charges
were quantum mechanical and that their appearance in classical mechanics
was exotic. This feeling is evident in references \cite{T2,T3} where
a model for gravity in \emph{two} spacetime dimensions was proposed
which involved a classical central charge in the Hamiltonian generators,
and also in \cite{BH} for gravity in three-spacetime dimensions.
(Incidentally it follows from \cite{CHvD} that the two-dimensional
model turns out to be the asymptotic dynamics of the three-dimensional
one, given by the Liouville theory so the two classical central charges
are one and the same).%
}).

However in the case of three-dimensional gravity the situation is
not that simple for two reasons: (i) one is dealing with a local field
theory and therefore it is unnatural to add by hand an overall single
degree of freedom (the central charge), (ii) the symmetry is an asymptotic
symmetry, whose definition and implementation critically depends on
the boundary condition at large distances and on what is allowed to
vary there.

Therefore it is necessary in this case to have an improved classical
formalism in which what will become the central charge in the asymptotic
symmetry algebra is a dynamical variable to start with. In other words,
the step described as optional for the particle in footnote {[}18{]}
is now mandatory.

A simple mechanism exists in which the cosmological constant is treated
as a dynamical field rather than as a universal constant \cite{ANT}
(see also \cite{T4}). The field then becomes independent of the point
in spacetime by virtue of the equations of motion. To the knowledge
of the present authors no analogous simple mechanism is available
to achieve a similar goal for the gravitational constant. For this
reason we will consider the superselection rule just proved for, 
\[
c=\frac{3\ell}{2G},
\]
as a superselection rule for, 
\[
\Lambda=-\frac{1}{\ell^{2}},
\]
and \emph{fixed $G$.}

The implementation of the mechanism will proceed along lines similar
to the analysis performed for four spacetime dimensions with $\Lambda<0$
in \cite{HT-1}. In that case, the finite dimensional $so(3,2)$ algebra
acquires an additional generator corresponding to the zero mode of
the field $\Lambda$, which commutes with $so(3,2)$, but no central
charge appears.

To make $\Lambda$ dynamical one introduces a two-form abelian gauge
potential $A$ so that the action reads, 
\begin{align}
I & =\frac{1}{16\pi G}\int d^{3}x\sqrt{-g}\left(R-2\Lambda_{0}\right)\nonumber \\
 & -\frac{1}{12}\int d^{3}x\sqrt{-g}F_{\mu\nu\lambda}F^{\mu\nu\lambda},\label{eq:action_grav+gauge}
\end{align}
where $F=dA$ is the gauge invariant field strength of the 2-form.

The general solution of the field equation associated to the abelian
gauge field can be written as, 
\begin{equation}
\left.F^{\mu\nu\lambda}\right|_{\text{on-shell}}=\left(-g\right)^{-1/2}E\epsilon^{\mu\nu\lambda},\label{eq:F-on-shell}
\end{equation}
where $E$ is an integration constant. Inserting (\ref{eq:F-on-shell})
into the equation for the gravitational field yields the Einstein
equations, 
\[
G_{\mu\nu}+\Lambda g_{\mu\nu}=0,
\]
with, 
\begin{equation}
\Lambda=\Lambda_{0}+4\pi GE^{2}.\label{eq:LambaE}
\end{equation}
The Anti-de Sitter case which is of interest in the present work,
is obtained by demanding that the ``bare'' cosmological constant
$\Lambda_{0}$ be negative, and such that the bound, 
\[
\Lambda_{0}+4\pi GE^{2}<0,
\]
on $E$ holds. One introduces $\Lambda_{0}$ in order to have a physical
Lagrangian with positive energy density for the dynamical $p$-form
gauge field.

The next step is to pass to the Hamiltonian form of the action (\ref{eq:action_grav+gauge})
and to give boundary conditions at spacelike infinity for both the
gravitational variables and the two-form gauge field which lead to
well defined surface integrals as symmetry generators \cite{RT}.
Since the cosmological constant has become a dynamical variable the
central charge will automatically arise as a symmetry generator. As
shown in the Appendix, after this is done, explicit forms for the
Witt algebra generators $L_{n}$ as surface integrals emerge and the
Poisson bracket algebra (\ref{eq:Vir}) is established. 

To conclude, a crucial step in the above derivation should be emphasized
and commented upon: it has been assumed that, upon passing to quantum
mechanics, the algebra (\ref{eq:Vir}) becomes (\ref{eq:Vir1}), i.e.,
that its form remains unchanged. We do not know how to prove this
assumption, but there is one remarkable fact that could be interpreted
as supporting its validity. It is the observation \cite{Srominger}
that, if taken literally, way beyond the scope of its derivation,
the algebra (\ref{eq:Vir1}) with the eigenvalue (\ref{eq:c}) reproduces
the standard formula for the black hole entropy in the semiclassical
limit %
\footnote{One would expect quantum mechanical corrections of the form $c=\left(3\ell/2G\right)\left(1+f\left(G\hbar/\ell\right)\right)$,
where $f\left(x\right)$ vanishes in the classical limit $x\rightarrow0$.
The assertion that there is a superselection rule for $\ell$ remains
valid because for fixed $G$ (and $\hbar$), $c$ is just a function
of $\ell$.%
}. Since that formula itself has also been obtained by applying a path
integral expression for the entropy beyond the domain of validity
of its derivation, one finds himself watching the consistency of two
mysteries. This would appear strange enough to be of significance.
\begin{acknowledgments}
The Centro de Estudios Científicos (CECs) is funded by the Chilean
Government through the Centers of Excellence Base Financing Program
of Conicyt. One of the authors (C.B.) wishes to thank the Alexander
von Humboldt Foundation for a Humboldt Research Award, while the other
(A.P.) gratefully acknowledges support through Fondecyt grant Nº 11130262. 
\end{acknowledgments}

\appendix

\section{Asymptotic symmetry algebra with a dynamical cosmological constant}

The Hamiltonian corresponding to the action (\ref{eq:action_grav+gauge})
in the main text has the form, 
\[
H=\int d^{2}x\left[N^{\perp}\mathcal{H}_{\perp}+N^{i}\mathcal{H}_{i}+A_{ti}\mathcal{G}^{i}\right]+Q\left[N^{\perp},N^{i},A_{ti}\right],
\]
with, 
\begin{eqnarray*}
\mathcal{H}^{\perp} & = & \frac{16\pi G}{\sqrt{\gamma}}\left[\pi^{ij}\pi_{ij}-\pi^{2}\right]-\frac{1}{16\pi G}\sqrt{\gamma}{}^{\left(2\right)}\left(R-2\Lambda_{0}\right)\\
 &  & +\frac{1}{\sqrt{\gamma}}P_{ij}P^{ij},\\
\mathcal{H}_{i} & = & -2\nabla_{j}\pi_{i}^{\; j},\\
\mathcal{G}^{i} & = & 2\partial_{j}P^{ij}.
\end{eqnarray*}
Here $\gamma_{ij}$ is the metric of the spatial section, $\pi^{ij}$
its conjugate momentum, and $P^{ij}$ the conjugate momentum to the
spatial 2-form gauge potential $A_{ij}$. The functions $\mathcal{H}_{\perp},\mathcal{H}_{i},\mathcal{G}^{i}$
are the constraint-generators of surface deformations and gauge transformations
of the 2-form gauge potential respectively. The lapse $N^{\perp}$,
the shift $N^{i}$ and the component $A_{ti}$ are the Lagrange multipliers
that parametrize the transformation generated by the corresponding
constraints. The surface integral $Q\left[N^{\perp},N^{i},A_{ti}\right]$
is the boundary term necessary to make the Hamiltonian a well defined
functional \cite{RT}. It generates the asymptotic symmetries.

The variation of the boundary term is, 
\begin{equation}
\delta Q\left[N^{\perp},N^{i},\lambda_{i}\right]=\delta Q_{g}\left[N^{\perp},N^{i}\right]+\int d\phi\left[2A_{t\phi}\delta P^{\rho\phi}\right],\label{eq:bdry term}
\end{equation}
where $\delta Q_{g}\left[\epsilon^{\perp},\epsilon^{i}\right]$ is
the usual boundary term of the pure gravitational field, given by,
\begin{align*}
\delta Q_{g}\left[N^{\perp},N^{i}\right] & =\int ds_{l}\left[\frac{1}{16\pi G}N^{\perp}G^{ijkl}\nabla_{k}\delta\gamma_{ij}\right.\\
 & -\frac{1}{16\pi G}\left(\nabla_{k}N^{\perp}\right)G^{ijkl}\delta\gamma_{ij}+2N^{i}\gamma_{ki}\delta\pi^{kl}\\
 & +\left.\left(2N^{k}\pi^{jl}-N^{l}\pi^{jk}\right)\delta\gamma_{jk}\right],
\end{align*}
with $G^{ijkl}=\frac{1}{2}\sqrt{\gamma}\left(\gamma^{ik}\gamma^{jl}+\gamma^{il}\gamma^{jk}-2\gamma^{ij}\gamma^{kl}\right)$.
The boundary term (\ref{eq:bdry term}) becomes an ``exact variation'',
i.e., the variation symbol $\delta$ can be taken outside the integral,
once boundary conditions are imposed to that effect.

\subsection{Asymptotic conditions}

In order to construct a consistent set of asymptotic conditions for
the coupled system of the gravitational field with the 2-form gauge
potential, we will proceed along the same lines of ref. \cite{HT-1}.
The key new element is that we now need to be able to vary the cosmological
radius $\ell$ through variation of $E$ in (\ref{eq:LambaE}). Therefore
it is natural to employ radial and time coordinates which are made
dimensionless through division by $\ell$. This choice proves indeed
useful because it facilitates the analysis of the surface terms arising
in the variation of the Hamiltonian. In terms of the dimensionless
$t$ and $\rho$, the fall-off conditions of ref. \cite{BH} read,
\begin{align}
\gamma_{\rho\rho} & =\ell^{2}\left[\frac{1}{\rho^{2}}+\frac{f_{\rho\rho}\left(\phi\right)}{\rho^{4}}+O\left(\rho^{-5}\right)\right],\nonumber \\
\gamma_{\rho\phi} & =\ell^{2}\left[\frac{f_{\rho\phi}\left(\phi\right)}{\rho^{3}}+O\left(\rho^{-4}\right)\right],\label{eq:asymp1}\\
\gamma_{\phi\phi} & =\ell^{2}\left[\rho^{2}+f_{\phi\phi}\left(\phi\right)+O\left(\rho^{-1}\right)\right],\nonumber 
\end{align}
and, 
\begin{align}
\pi^{\rho\rho} & =O\left(\rho^{-1}\right),\nonumber \\
\pi^{\rho\phi} & =\frac{1}{\ell^{2}}\left[\frac{p^{\rho\phi}\left(\phi\right)}{\rho^{2}}+O\left(\rho^{-4}\right)\right],\label{eq:asymp6}\\
\pi^{\phi\phi} & =O\left(\rho^{-5}\right).\nonumber 
\end{align}
In addition, we will demand the asymptotic behavior of the spatial
component of the gauge field and its canonical momentum to be, 
\begin{align}
A_{\rho\phi} & =O\left(\rho^{-1}\right),\nonumber \\
P^{\rho\phi} & =\frac{E}{2}+O\left(\rho^{-5}\right).\label{eq:asympt8}
\end{align}
The asymptotic behavior of the Lagrange multipliers $N^{\perp},N^{i}$
and $\lambda_{j}$ is in turn determined by requiring that the fall-off
(\ref{eq:asymp1})-(\ref{eq:asympt8}) be preserved under the symmetry
transformations. Denoting with a prime the derivative with respect
to the angle $\phi$, this gives, 
\begin{align}
N^{\perp} & =\frac{\ell}{2}\rho\left(T^{+}\left(\phi\right)-T^{-}\left(\phi\right)\right)+O\left(\rho^{-1}\right),\nonumber \\
N^{\rho} & =-\frac{\rho}{2}\left(T^{+'}\left(\phi\right)+T^{-'}\left(\phi\right)\right)+O\left(\rho^{-1}\right),\label{eq:param1}\\
N^{\phi} & =\frac{1}{2}\left(T^{+}\left(\phi\right)+T^{-}\left(\phi\right)\right)+O\left(\rho^{-1}\right),\nonumber 
\end{align}
just as in \cite{BH}, whereas for the gauge field multipliers one
finds, 
\begin{align}
A_{t\rho} & =O\left(\rho^{-1}\right),\nonumber \\
A_{t\phi} & =-\frac{\ell^{3}}{4}E\left[\rho^{2}\left(T^{+}\left(\phi\right)-T^{-}\left(\phi\right)\right)^{\phantom{^{^{^ {}}}}}\right.\nonumber \\
 & \left.-\frac{1}{2}\left(T^{+}\left(\phi\right)-T^{-}\left(\phi\right)\right)f_{\rho\rho}-12\lambda+O\left(\rho^{-1}\right)\right].\label{eq:lambdaphi}
\end{align}
As it also happens in the four dimensional case \cite{HT-1}, eq.
(\ref{eq:lambdaphi}) shows that the Lagrange multiplier $A_{t\phi}$
explicitly depends on the parameters $T^{\pm}$ that determine the
asymptotic displacements in spacetime. This delicate interplay has
two crucial effects: (i) The form of the $\rho^{2}$ piece cancels
a divergence in the charges coming from the pure gravitational part,
(ii) The form of the order $\rho^{0}$ piece makes (\ref{eq:bdry term})
an ``exact differential'', which guarantees that the charges can
be integrated.

\subsection{Asymptotic symmetries and global charges.\label{sub:Asymptotic-symmetries-and}}

Imposing the boundary conditions (\ref{eq:asymp1})-(\ref{eq:asympt8}),
and (\ref{eq:param1})-(\ref{eq:lambdaphi}) we obtain for the surface
integral which generates the asymptotic symmetries, 
\begin{align*}
Q\left[T^{+},T^{-},\eta\right] & =\int d\phi\left\{ T^{+}\left(\phi\right)L^{+}\left(\phi\right)-T^{-}\left(\phi\right)L^{-}\left(\phi\right)\right\} \\
 & +\lambda c,
\end{align*}
where, 
\begin{align*}
L^{\pm}\left(\phi\right) & =\frac{\ell}{32\pi G}\left(f_{\rho\rho}\left(\phi\right)+2f_{\phi\phi}\left(\phi\right)\right)\pm p^{\rho\phi}\left(\phi\right),\\
c & =\frac{3\ell}{2G}.
\end{align*}
We see that indeed the central charge has become the generator of
a global symmetry transformation (improper gauge transformation in
the terminology of \cite{CT}). A displacement generated by $\mathcal{Q}$
of magnitude $\lambda$ corresponds to a global gauge transformation
with, 
\[
A_{t\phi}=3\ell^{3}E\lambda,
\]
at infinity.

To determine the asymptotic symmetry algebra, one expresses the commutator
of two asymptotic symmetry transformations in terms of their individual
asymptotic parts. If we denote collectively by $\xi=\left(\xi_{+},\xi_{-},\xi_{\lambda}\right)$
the parameter of an asymptotic symmetry transformation, the transformation
law of $L^{\pm}$ and $c$ is given by, 
\begin{align*}
\delta_{\xi}L^{\pm} & =\xi^{\pm}L^{\pm'}+2L^{\pm}\xi^{\pm'}-\frac{c}{24\pi}T^{\pm'''},\\
\delta_{\xi}c & =0.
\end{align*}
One then finds for the parameter $\zeta$ of the commutator $\delta_{\zeta}=\delta_{\eta}\delta_{\xi}-\delta_{\xi}\delta_{\eta}$,
of two transformations with parameters $\xi$ and $\eta$, 
\begin{align*}
\zeta_{\pm}\left(\phi\right) & =\eta_{\pm}\left(\phi\right)\xi_{\pm}^{'}\left(\phi\right)-\xi_{\pm}\left(\phi\right)\eta_{\pm}^{'}\left(\phi\right),\\
\zeta_{\lambda} & =0.
\end{align*}
When expressed in terms of Fourier modes these equations read, 
\begin{align*}
\zeta_{\pm}^{s} & =\left(m-n\right)\delta_{m+n}^{s}\xi_{\pm}^{m}\eta_{\pm}^{n},\\
\zeta_{\lambda} & =0,
\end{align*}
which characterize two copies of the Witt algebra. There is no central
charge when the algebra acts on spacetime through deformations of
hypersurfaces. The central extension appears when one realizes the
asymptotic symmetries in terms of Poisson brackets, and the Witt algebra
becomes then the Virasoro algebra.

\subsection{Canonical realization of the asymptotic symmetry algebra.}

The realization of the asymptotic symmetry algebra in terms of Poisson
brackets is obtained from the definition, 
\[
\left\{ Q\left[\xi\right],Q\left[\eta\right]\right\} =\delta_{\eta}Q\left[\xi\right],
\]
through direct evaluation of the asymptotic part of the variation
on the right-hand side. Expanding in Fourier modes $L^{\pm}\left(\phi\right)=\frac{1}{2\pi}\sum_{n}L_{n}^{\pm}e^{in\phi}$
one finds, 
\begin{align*}
i\left\{ L_{m}^{\pm},L_{n}^{\pm}\right\}  & =\left(m-n\right)L_{m+n}^{\pm}+\frac{c}{12}m^{3}\delta_{m+n,0},\\
i\left\{ L_{m}^{\pm},c\right\}  & =0,\\
i\left\{ c,c\right\}  & =0,
\end{align*}
as just anticipated. (Replacing $L_{0}$ in the above equations by
$L_{0}-c/24$ yields equations (\ref{eq:Vir}) in the main text).

\end{document}